\documentclass[10pt,journal]{IEEEtran}

\IEEEoverridecommandlockouts

\usepackage{amsmath,amsfonts,amsthm,amssymb}
\usepackage[font=scriptsize,caption=false,labelsep=space]{subfig}
\usepackage{graphicx,float}
\usepackage{graphics}
\usepackage{mathtools}
\usepackage{verbatim}
\usepackage{xcolor}
\usepackage{url}
\usepackage{multirow}
\usepackage{cite}

\usepackage{balance}
\usepackage[none]{hyphenat}

\definecolor{darkgreen}{rgb}{0.01, 0.75, 0.24}

\begin{document}
%
\title{
Non-Contact Vital Signs Detection with \\UAV-Borne Radars}

%

\author{Yu~Rong,~\IEEEmembership{Member~IEEE,}
	Richard~M. Gutierrez,~\IEEEmembership{Member~IEEE,} 
	Kumar~Vijay Mishra,~\IEEEmembership{Senior Member~IEEE,}
	and~Daniel~W. Bliss,~\IEEEmembership{Fellow~IEEE}
	\thanks{Y. R. and D. W. B. are with Center for Wireless Information Systems and Computational Architectures (WISCA) at the School of Electrical, Computer and Energy Engineering, Arizona State University, Tempe, AZ 85281 USA, e-mail: yrong5@asu.edu.}
	\thanks{R. M. G. is with the General Dynamics Mission Systems, Scottsdale, AZ 85257 USA, e-mail: richardmgutierrez@asu.edu.}
	\thanks{K. V. M. is with the United States Army Research Laboratory, Adelphi, MD 20783 USA, e-mail: kumarvijay-mishra@uiowa.edu.}
}

\markboth{Journal of \LaTeX\ Class Files}%
{ }

\maketitle

\begin{abstract}
 Airborne radar carried on-board unmanned aerial vehicles (UAV) is serving as the harbinger of new remote sensing applications for security and rescue in inclement environments. The mobility and agility of UAVs along with intelligent on-board sensors (cameras, acoustics, and radar) are more effective during the early stages of disaster response. 
 The ability of radars to penetrate through objects and operate during low visibility conditions enables detection of occluded human subjects on and under debris when other sensing modalities fail. Recently, radars have been deployed on UAVs to measure minute human physiological parameters such as respiratory and heart rates while sensing through clothing and building materials. 
 Aggregating radar measurements with the information from other sensors is broadening the applications of drones in life-critical situations. 
 Signal processing techniques are critical in enabling UAV-borne radars for human vital sign detection (VSD) in multiple operation modes. Novel radar configurations such as in a UAV swarm and tethered UAVs are required to facilitate multi-tasking and high endurance, respectively. 
 This paper provides an overview of recent advances in UAV-borne VSD with a focus on deployment modes and processing methods. 
\end{abstract}


%
\IEEEpeerreviewmaketitle

\section{Introduction}
\label{sec:intro}
\IEEEPARstart{U}{nmanned} aerial vehicles (UAV), also referred as drones, are emerging as valuable tools for surveillance, medical assistance, consumer goods deliver, and policing \cite{zhao2019uav}. In particular, commercial drones are now deployed for disaster response including damage assessment, map updates, and search-and-rescue operations (SRO) for stranded victims \cite{erdelj2017help}. 
In this paper, we focus on this latter application of UAV-based physiological \textit{vital sign detection} (VSD) of victims.

When considering emergency response in a crisis, early determination of location and status of victims is a priority. 
In this context, drones exhibit enhanced mobility, accessibility, agility, and height advantages over humans, animals, and robots on the ground. A UAV is more efficient and less error-prone than human rescue teams and trained service dogs under extreme environments, shortage of time, and difficult physical conditions. Compared to ground robots, UAVs do not require invasive equipment and their mobility is not considerably challenged in off-road and complex terrains with mudslides, avalanches, flooding. 


The performance of SRO UAVs is significantly improved by equipping them with intelligent sensors that are powered by computer vision and machine learning techniques. These sensors comprise specialized optical devices in addition to sensors for flight control, stabilization, and communications. The optical sensors are task-specific and cater for functions such as super high-resolution (SHR) red-green-blue (RGB) cameras for detailed vision inspection during daytime; infrared thermal cameras for creating a pseudocolor heat map of the scene from temperature measurements; and time-of-flight distance sensors (depth sensors or lidar). In most cases, fusing measurements from depth sensors and the vision cameras provides an accurate three-dimensional (3-D) map. 

\begin{figure}[t]
	\centering
	\includegraphics[width=\linewidth]{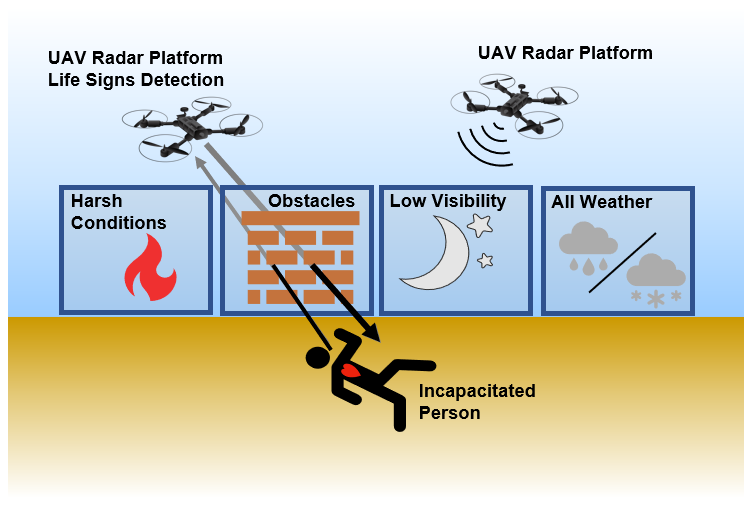}
	\caption{A radar mounted on UAV (or UAV swarm) detects living humans on the ground by extracting their vital physiological measurements from the received echoes. Unlike optical systems, radar signals leverage the see-through capability at microwave in harsh weather, low visibility condition or when the victims are covered under debris.} \label{fig::intro application}
	\label{fig:uav_vsd}
\end{figure}

The aforementioned non-contact sensors are also suitable for measurements of physiological signs of any living human on the ground from a UAV platform (Fig.~\ref{fig:uav_vsd}). Although camera systems are favorable for line-of-sight object detection and lidar systems come with high spatial resolution, these sensing modalities are limited by debris and building material that cover or occlude the victim at the disaster site. On the other hand, microwave signals at appropriate radar frequencies not only penetrate through such obstacles but also perform well in unfavorable conditions such as harsh weather and low visibility. In this paper, we focus on radar-based VSD from UAV platforms. 

We first provide a synopsis of prior works on UAV-based VSD while comparing the radar and optical sensors. We follow this with an overview of design considerations of UAV-borne radars for VSD. We then describe major deployment modes and results from experimental demonstrations from existing prototypes. Finally, we identify major technological challenges and highlight future research directions.

\section{UAV-based VSD: State-of-the-Art}
\label{sec:sota}
\begin{table*}
  \caption{Summary of the state-of-the-art in UAV-based non-contact VSD. }
  \label{tbl:review}
  \includegraphics[width=1.0\textwidth]{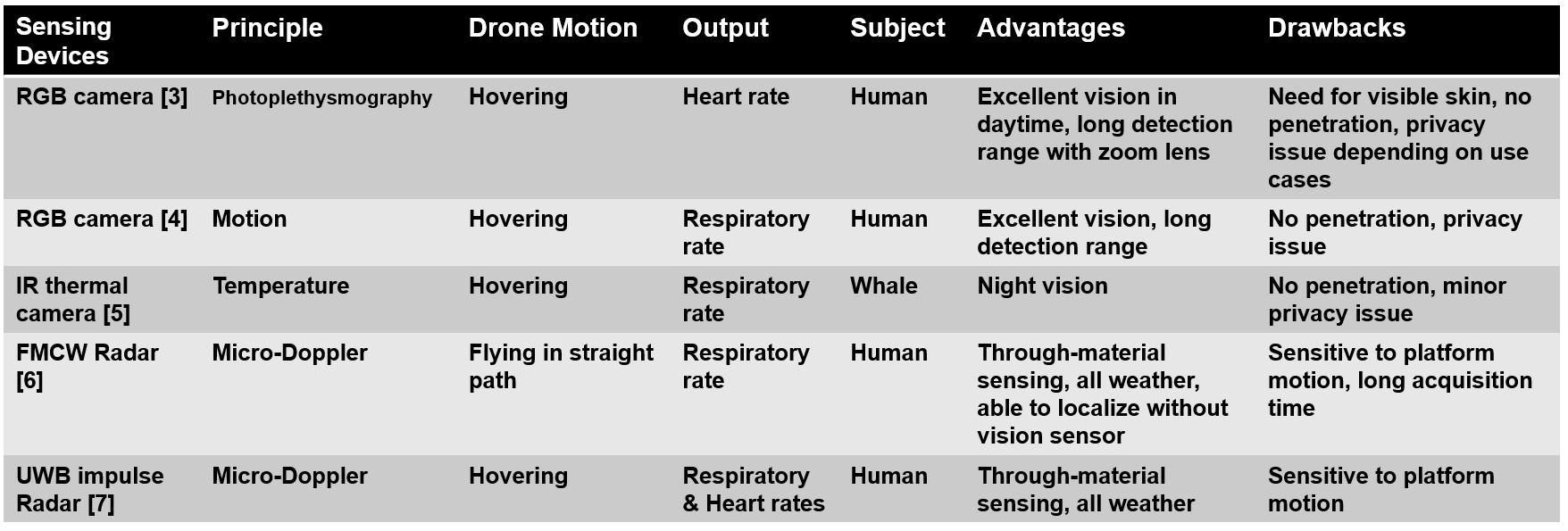}
\end{table*}
A radar transmitter emits a periodic train of known narrowband pulse which interacts with the target-of-interest. The resulting reflections bouncing off from the targets are received by the radar as superimposed, attenuated, time-shifted, and frequency-modulated (or Doppler-shifted) replicas of the transmit signal. The time delay and frequency modulation in the received signal are directly proportional to the unknown range and velocity of the target, respectively. By estimating these unknown parameters in the received signal, the radar determines the physiological parameters such as respiratory and heart rates of the human subject.

The radar and optical frequencies, including that of visible and invisible light, occupy distinct positions in the electromagnetic spectrum. Whereas radar carrier wavelength for UAV VSD applications ranges from fraction of a meter to a few millimeters, the same for optical systems goes from approximately 380 nanometers to a millimeter. As a result, the physiological phenomenology and hardware differ considerably for radar and optical sensors. In case of radar, the received signal containing the human physiological signature exhibits the \textit{micro-Doppler} effect arising from the micro-motions of body parts such as the lungs and heart. For example, breathing results in the displacement of a few millimeters and heartbeat that of much less then a millimeter. The phase of the received signal is sensitive to these tiny motions and extraction of the Doppler shift embedded in the phase provides a good estimation of the vital signs. 
Smaller wavelengths (or higher frequencies) yield better phase sensitivity but at the cost of shorter sensing distances and lower penetration capability. This system design trade-off must be considered while employing UAV radars for VSD. Table~\ref{tbl:review} summarizes these differences between radar and optical approaches as reported in major prior works on UAV-based VSD \cite{al2018remote,perera2020detection,horton2019doctor,yan2018vital,rong2021UAV} .

 Stand-alone small-scale radars have been available for measuring breathing and heart rates of human based on Doppler effect since 1970s \cite{lin1975noninvasive}. This was followed with advancements in signal processing techniques, hardware, and antenna designs to enable radar as a potential non-contact remote sensing technology for health monitoring. Recently, through-wall multiple heartbeats detection was demonstrated with wideband radar \cite{rong2019smart}. Using multiple-input multiple-output (MIMO) radar imaging, precise pulse wave have also been reconstructed \cite{rong2019radar}. 

It is worth mentioning here another widely used non-contact VSD technique, i.e., remote imaging photoplethysmography (RIPPG) \cite{feng2014motion}. Instead of exploiting the Doppler effect of moving body parts as in radars, the RIPPG employs optics to detect minute color shifts caused by blood circulation in the naked skin of facial areas, hands, and feet to estimate the vital signs. This variation in radiant intensity of human skin with pulses of blood and motions is not discernible to the human eye but easily captured in a sequence of video frames. However, RIPPG is not robust against dynamic changes in visibility conditions and skin tones. Other related techniques include infra-red (IR) thermography, which extracts breathing signatures by evaluating the temperature changes of the human nostril \cite{horton2019doctor}. Depth sensors have also been used to directly measure physical displacement of the chest to estimate breathing rate. In general, this outperforms the efforts to extract chest motion by tracking image pixel value change \cite{rong2020DepthSensor}. Nearly all of these other sensors are incapable of penetration through obstacles in disaster areas. This makes radar very attractive and beneficial for UAV-based VSD.

\begin{figure*}[t]
	\centering
	\includegraphics[width=1\textwidth]{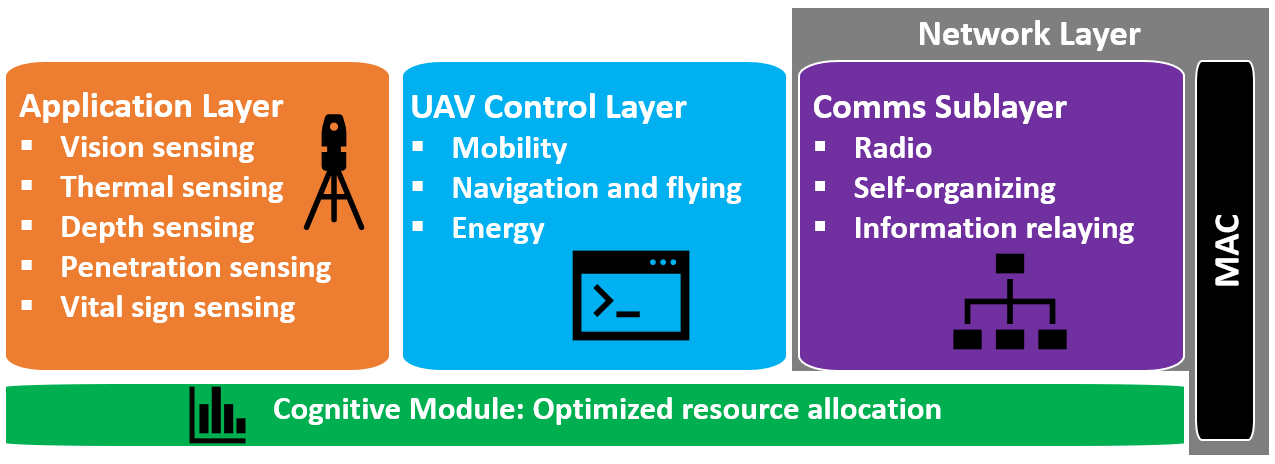}
	\caption{The layered organization of modules for VSD UAV systems modified from \cite{camara2014cavalry}, including application layer for autonomously coordinating tasks; control layer for self-navigation and flying; network layer for communications (comms) and connectivity; and cognitive module for decision making and resource allocation across layers.} 
	\label{fig::System modules}
\end{figure*}

In practice, a successful disaster response and SRO involves path planning, resource allocation, human localization, and life-signs detection; each of these tasks is accomplished using multiple sensing modalities. The UAV radar measurements usually complement the high-resolution information obtained from vision sensors. 
In \cite{kishk2020aerial}, a UAV-based communications sensor for used as a mobile base station to enhance coverage and capacity of cellular networks by leveraging the inherent relocation flexibility of the drone and a higher probability of establishing a line-of-sight (LoS) link with the ground mobile users. To accurately detect open water swimmers, UAV-based video identification has been used. 
In the context of cetacean conservation, \cite{horton2019doctor} proposed UAV IR thermography for monitoring vital signs of free-ranging whales. The UAV radar-based VSD is a relatively new area and, hereafter, we focus on design and processing aspects of only this technique.



\section{Design and Prototypes}
\label{sec:design}

Most commercial drones have a layered organization of hardware modules (Fig.~\ref{fig::System modules}). Connected to the physical layer sensors is the network layer with Medium Access Control (MAC), which provides the network connectivity to all other modules. Within the same layer is the communications sub-layer that comprises radio management for power control and optimization of communications with other drones within reach; self-organizing network module for exchanging messages and coordinating with nearby drones; and information relay for receiving data from other drones and forwarding them to the next drone, or broadcast it until the availability of destination and other drones. 

Connected to all layers is a cognitive module that provides generic artificial intelligence (AI) algorithms to help with decision making. A UAV control layer is responsible for the motion and routing. Its mobility management module performs path planning while also considering the objectives and probable actions of neighboring drones. This plan is then implemented by a separate navigation and flying control module. The layer also has energy management to keep track of the left-over battery power. 

The application layer is task-specific and dependent on the available sensors. For the LoS scenario, VSD involves fusion of vision and radar information. The color/IR cameras are used first to identify the contours and body features of a possible human subject through computer vision algorithms. Then, the region-of-interest (chest area) is identified and, through a vision-aided mechanical/electronic steering, the radar beam illuminates the chest for vital sign extraction. Note that the radar- and vision-based biometric data are used together for cross validation and sensor fusion to enhance the robustness of measurements. However, in a non-LoS situation such as an obstructed victim, penetrating signals from only radar are useful. The UAV uses an on-board communications system to identify hotspot areas where victims might be stranded. For example, This is helpful in identifying subjects with mobile phones. Even though such victims may not be able to make a call because of injury or the breakdown of communications infrastructure, it is possible to use the mobile GPS and handshake signals to locate them. Then UAV scans the focused area using a radar and detects living humans based on valid micro-Doppler signatures.

\subsection{Design considerations} \label{section::design considerations}
The installation of a radar on-board a UAV should be such that it does not hinder the common functions of the platform. Specific to VSD, the following factors determine the design. 
    \subsubsection{Loading capability} The payload is the weight that a drone carries. The greater payload a drone takes, the more flexible it is to supplement the system with extra cameras, sensors, packages for delivery, and additional technology to adapt to specific needs. However, the flight time is reduced when carrying more weight because the extra power required drains the battery sooner.
    
    \subsubsection{Flight time} 
    Longer flight times are required for extensively surveying a given area. Further, emergency situations typically require frequent monitoring of sites. The limited battery power constrains the flight times to usually less than one hour in most commercial UAVs. The overall design of the drone also affects the flying duration. For instance, Parrot Disco is capable of a continuous flight time of 45 minutes but a similar weight drone DJI Mavic Pro operates continuously only for 30 minutes. The fixed-wing, single-rotary blade in the former is aerodynamically better (less thrust) than the multi-rotor quadcopter design of the latter. 
    
    \subsubsection{Radar placement} Usually, optical flow camera is installed in the nose of the drone for unobstructed field of view. The same applies to the installation of radar sensors, which should also not interfere with the existing cameras and sensors. An alternative deployment location is the underbelly of the drone. But it should not block the view of ultrasound sensor that is normally located under the belly, vertically facing down for determining the height from the ground and detecting the objects directly underneath the drone. The top of the drone usually has an altimeter, a 9-axis inertial measurement unit (IMU), and a global positioning system (GPS) in protective casings. These do not compete with the radar for deployment location.
    
    \subsubsection{Platform motion compensation (PMC)} The drone-based sensing systems should be able to handle motion noise during flight and still provide precise measurement for applications like aerial surveying, radar imaging, and human physiological signal measurements. There are also perturbations in platform motion while hovering. Strategically, the platform motion is mitigated by mechanical stabilization, correction using digital signal processing, and sensor fusion. For example, a drone-based FLIR imaging solution combines a high-resolution radiometric thermal imager and 4K resolution RGB camera to create accurate orthomosaics. Similarly, a lidar and RGB solution from DJI for aerial surveys integrates a Livox lidar, a high-accuracy IMU, and a color camera. Radar sensors may also be likewise integrated with cameras. Further, using IMU and real-time kinematic GPS, a radar-vision coupled multi-modality payload is installed on the gimbals of drones for emergency response, highly accurate 3D mapping, and VSD in complex environments. 

\subsection{Customized prototypes}
\begin{figure}[t]
	\centering
	\includegraphics[width=1.0\columnwidth]{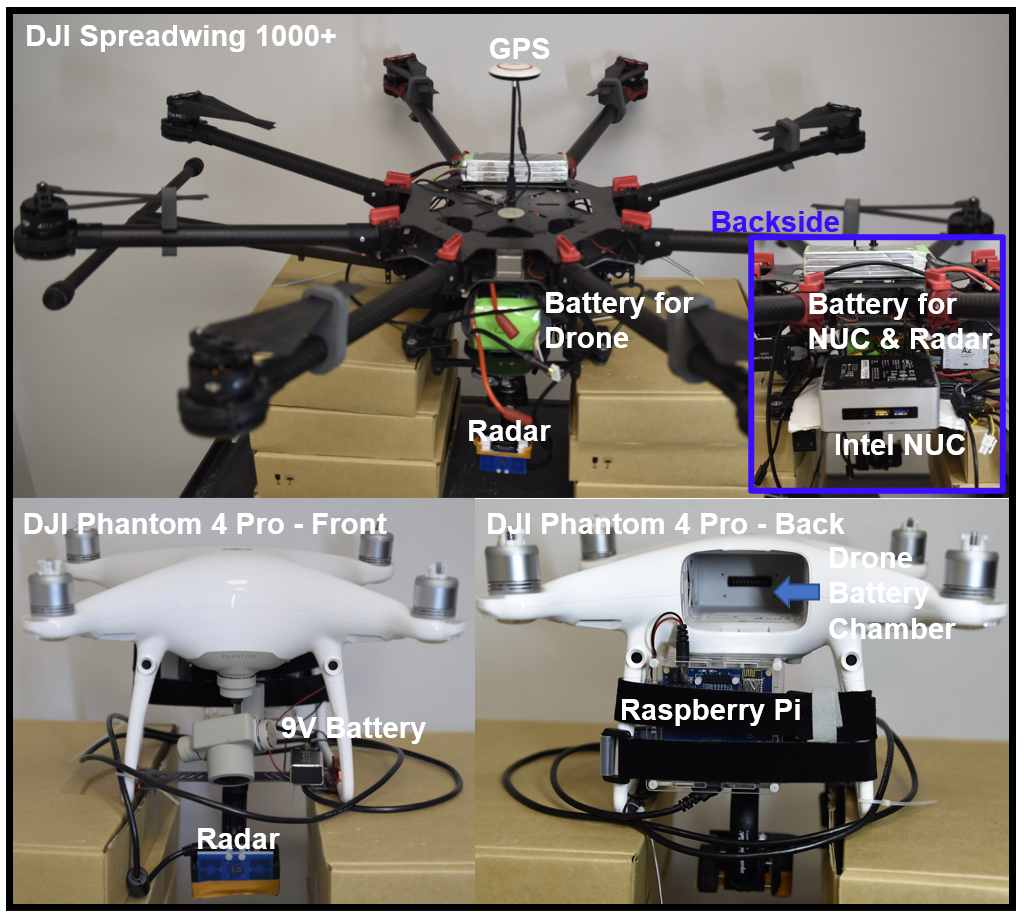}
	\caption{ASU UAV radar demonstration platforms. Top: DJI Phantom 4 Pro (quadcopter) uses Intel NUC mini PC. Bottom: DJI Spreadwing 1000+ (octocopter) employs Raspberry Pi.} \label{fig::UAV Systems}
\end{figure}

Center for Wireless Information Systems and Computational Architectures (WISCA) at Arizona State University has developed two UAV radar demonstration platforms (Fig. \ref{fig::UAV Systems}) assembled from off-the-shelf hardware components for demonstrating the proof-of-concept of UAV-based radar VSD. 
The flying carriers were DJI Phantom 4 Pro and DJI Spreadwing 1000+. The former is a quadcopter with 3 lbs weight, 35 cm diagonal length, flight time of 30 minutes and 3 lbs payload lift capability. The latter is a larger octocopter with 10 lbs weight, 104.5 cm diagonal length, 20 minutes flight time and 11 lbs lifting capability. The installed radar system consists of an ultra-wideband (UWB) impulse radar SoC (system-on-chip); a Raspberry Pi single-board computer (SBC) for controlling the radar and communicating with the ground personal computer (PC); and power sources for radar and SBC. In the next section, while discussing common deployment modes, we include demonstration of human VSD using these prototypes. These experiments were performed outdoors in Phoenix, AZ and followed the government regulatory requirements. 

\section{Deployment Modes} \label{Section::VSD examples}
\label{sec:modes}
The deployment configuration is critical to the quality of VSD with UAV-borne radars. The common processing modes employ radar interferometry, synthetic aperture radar (SAR), and through-material imaging. We differentiate these UAV radar methods by not only radar signal processing but also UAV maneuvers during the radar scanning period. 

\subsection{Interferometry mode}
\begin{figure*}[t]
	\centering
	\includegraphics[width=1.0\textwidth]{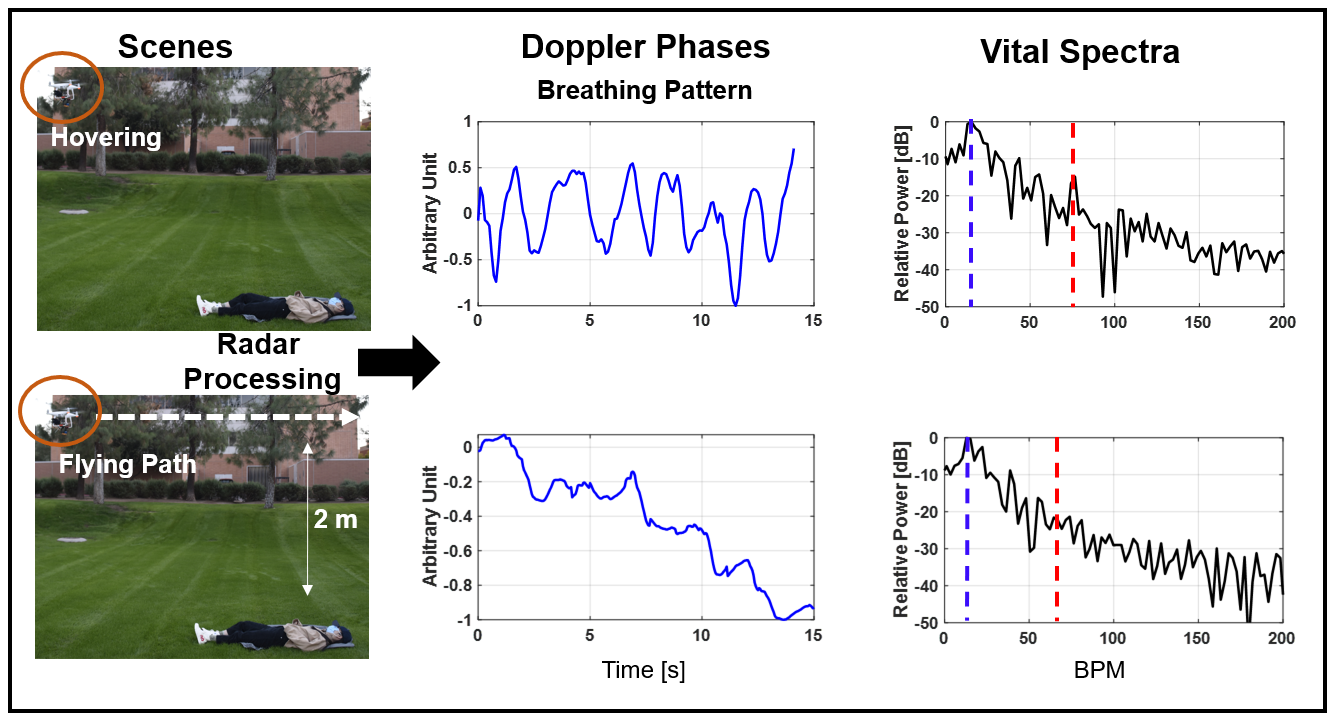}
	\caption{UAV-borne radar interferometry examples using two flight maneuvers: hovering and flying in a straight path. The breathing signal is detected in both cases. However, the heartbeat is detected only during hovering. The dashed blue and red lines denote breathing and heartbeat references; BPM stands for beats per minute. The UAV is enclosed within a brown circle in the images on left.} \label{fig::Interferometry}
\end{figure*} 
The interferometry mode implies that only micro-Doppler information is used in the radar processing. This mode discriminates humans from other non-living objects by strategically flying the UAV over certain flying routes. Since the distance between the radar and target constantly changes, periodic chest motion vibrations are extracted via micro-Doppler processing by combining data from all range profiles. Common flight maneuvers include linear straight path at constant altitude; circular path centered on the human target at constant altitude; and approaching the human target while changing only the height of the drone. Note that many commercial drones allow user to pre-program the flying route for more precise flying experience. For example, DJI software features \textit{Waypoints} function that allows customization of flying path, distance (as small as a few meters), and speed. 

Some representative interferometry results from WISCA prototypes are shown in Fig.~\ref{fig::Interferometry}. When the drone hovers right above the human chest area at close distances, both breathing and heartbeat signatures are easily identified. More specifically, breathing signal variations are retrieved and heart rate is estimated during this period. When the drone moves strategically around the target, the resulting measurement is noisy but a discernible micro-Doppler signature from regular chest vibration is still observed in spectral analyses.

\subsection{Vital-SAR imaging}
\begin{figure*}[t]
	\centering
	\includegraphics[width=1\textwidth]{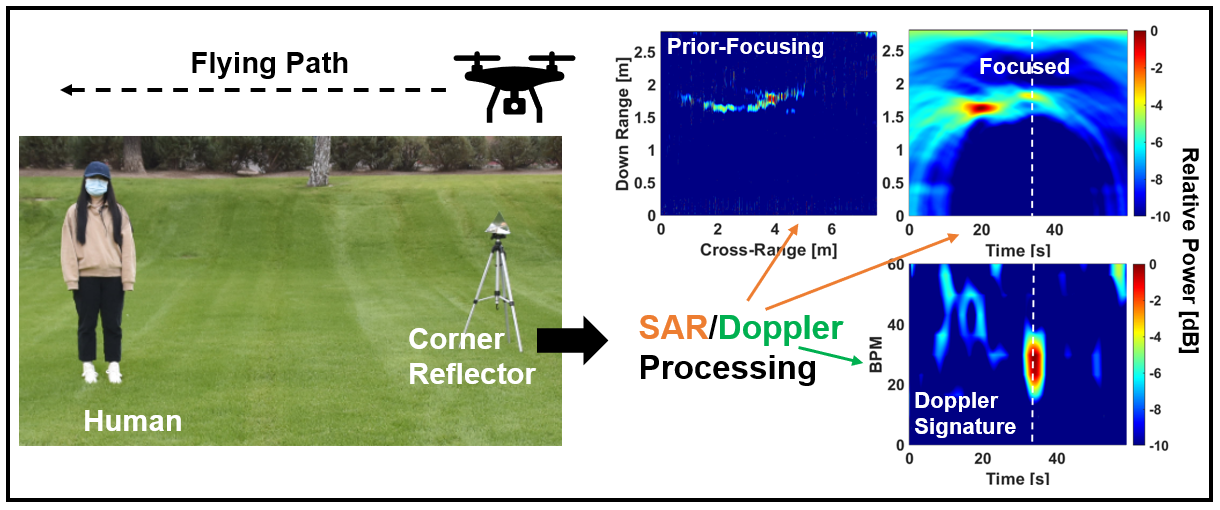}
	\caption{Illustration of vital-SAR imaging. The RGB image of the scene (left) shows a corner reflector and a human subject on the scene scanned by the UAV flying from right to the left. When the raw signal (upper middle) from SAR is focused (upper right), only the locations of both targets are revealed. The VSD processing (lower right) then identifies the human micro-Doppler (breathing).} \label{fig::VSAR}
\end{figure*} 
In the SAR mode, the radar employs an antenna array on the drone to image targets through several measurements while continuing the flight movement. This synthesizes a larger antenna aperture to achieve a finer spatial resolution than possible with the limited aperture of the on-board antenna. The SAR processing generates intensity map image that indicates the location of strong scatterers. In \cite{yan2018vital}, SAR-based VSD or \textit{vital-SAR imaging} was introduced. Here, SAR processing is combined with VSD to discriminate living humans from static clutter in the reconstructed SAR image. The SAR image formation and vital sign extraction themselves were accomplished separately by flying the drone twice over the same path.

The WISCA prototypes improve \cite{yan2018vital} by performing vital-SAR imaging in a single flying path and without using vision sensors. 
This setup comprises a, X-band UWB impulse radar with 1.5 GHz bandwidth installed at the bottom of the drone. During the flight, the radar always points in a downward direction perpendicular to the straight flight path. Note that, compared to the SHR camera, a radar SAR image provides limited and abstract vision content. But SAR clearly outperforms SHR in harsh weather, obstructions, and low visibility conditions. Figure~\ref{fig::VSAR} shows an example of vital-SAR imaging with WISCA prototype. The SAR image shows two strong energy clusters corresponding to a corner reflector and a human subject in the RGB image of the scene. The time-frequency representation of breathing signal then identifies the living human subject  from the corner reflector.  

\subsection{Through-material imaging}
\begin{figure*}[t]
	\centering
	\includegraphics[width=1\textwidth]{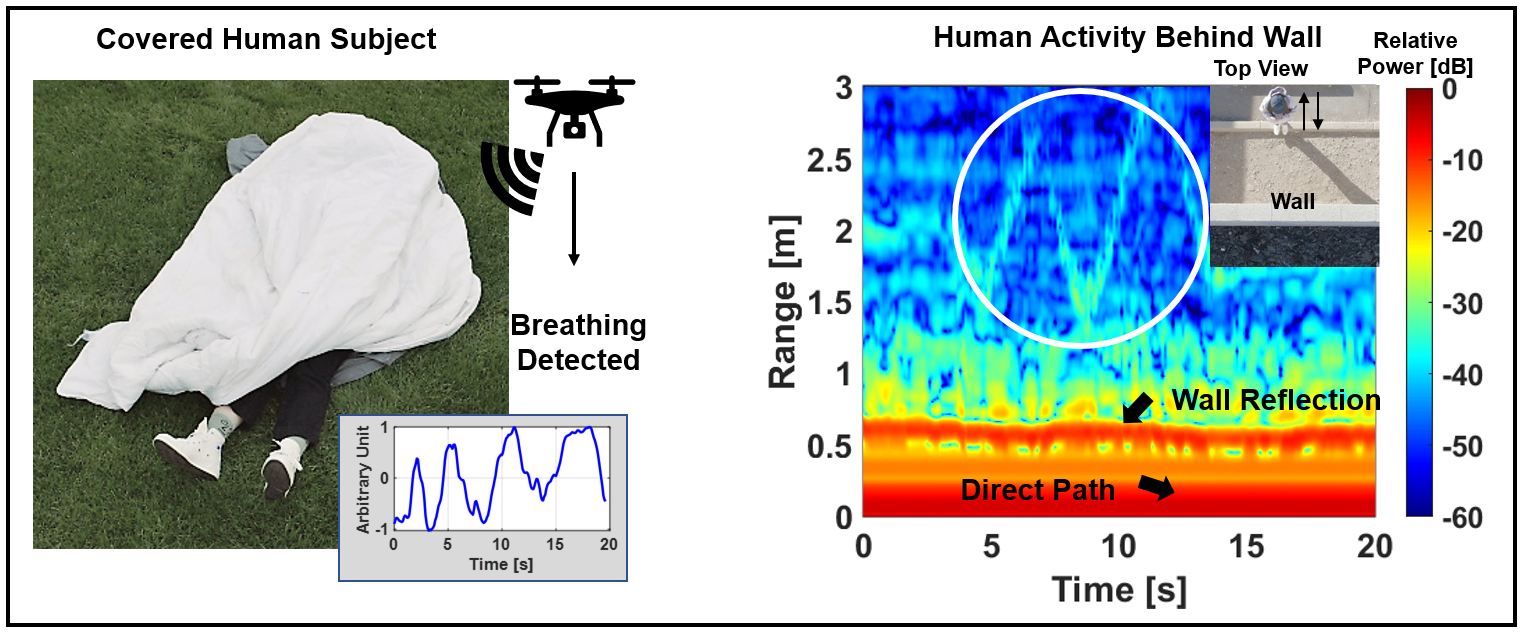}
	\caption{Through-material imaging for VSD. Left: The UAV radar extracts the breathing pattern (inset) of a human subject covered with bedding. Right: The same radar performs the through-wall imaging of a walking human (inset), whose trace is enclosed by the white circle.} \label{fig::See Through}
\end{figure*} 
As mentioned earlier, through-material imaging is the major advantage while using radar over optical sensors. Figure~\ref{fig::See Through} shows two examples of WISCA UAV radar in detecting vital signs of an obstructed subject. In the first example, the breathing pattern is recovered when the human subject was covered by bedding material; for a radar, such an obstruction causes insignificant signal attenuation. In the second scenario, the UAV-borne radar detects a human walking activity through a wall. Note that the locations of wall reflections slowly drift over time because of the continually moving UAV platform.

\section{Challenges and Future Directions}
\label{sec:chal}
Large coverage, quick response, constant monitoring, and precise surveying are major challenges for deploying UAVs in disaster response and similar situations for VSD. While we focused on radar processing, design, and deployment, several other important aspects are also currently under investigation. These include new regulatory policies, data integrity, privacy concerns, and reliability of UAVs with respect to accidents or malicious risks. Some emerging applications and considerations of UAV radar VSD are highlighted below.
    \subsection{VSD with UAV swarms}
   For quick coverage of a large area, a swarm of multiple UAVs is capable of coordinated scanning. A heterogeneous swarm with different types of drones provides multi-tasking ability to perform various missions such as providing temporary communications service, disaster site mapping, SRO, and medical material delivery. Each of these activities require specific competence, thus requiring more than one type of drones and sensor equipment. Multiple drones also perform collaboratively to accomplish a single task with greater efficiency and improved performance. For example, in \cite{alaee2019radar}, a distributed UAV radar array is investigated to provide a desired beampattern for communications and sensing. Such coordinated UAV swarm signal processing is critical for improving the current estimates of vital signs obtained via a single drone.
    
    \subsection{VSD with tethered airborne sensing platforms}
    Short flight and service times of most commercially available UAVs determine the need to frequently recharge. This temporarily leaves UAVs out of service. Besides the efforts on developing new propulsion technology, drone aerodynamic design, efficient communications, on-board processing, and solar-powered UAVs, the emerging technology of \textit{tethered UAV} (tUAV) addresses the limitations in flight times. The tUAVs are continuously powered and have been previously proposed to establish a reliable/efficient backhaul link to ground station for data transfer and communications services \cite{kishk2020aerial}. These multi-purpose tUAV platforms are feasible for long-term VSD services. 
    
    \subsection{VSD with signal-of-opportunity PMC}
    In the absence of wind, the measured platform motion of UAVs in hovering state is usually of the order of a few centimeters. This perturbation in the motion is significantly larger than a typical physical displacement of chest motion and heartbeat. The PMC techniques discussed in Section \ref{section::design considerations} exploited mechanical cancellation and sensor fusion. 
    When such methods are unavailable, differential detection using opportunistic signals may be used in through-wall VSD from a hovering UAV. Assuming the radar emits wideband signals, the reflections from the wall and the human subjects are separated in the range profile. The backscattered energy from the wall contains only the platform motion while that from the human has both platform motion and vital signs. Through advanced signal processing of mixing the two signals, the VSD signal is obtained with correction for the platform motion. 
    
    \subsection{UAV VSD for pandemic mitigation}
\begin{figure}[t]
	\centering
	\includegraphics[width=\linewidth]{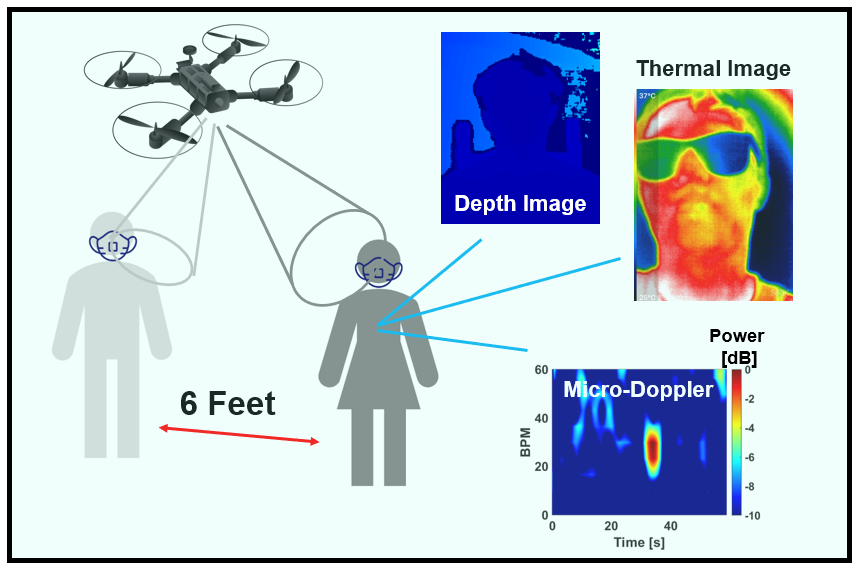}
	\caption{Various VSD sensors installed on UAVs  aid in patient monitoring during pandemics such as COVID-19. Upper middle: Coupled with RGB images, the depth sensors provide remote 3-D information about adherence to social distancing. Upper right: Thermal sensors dispense body temperature images through a non-contact interaction with the patient. Bottom right: UAV radars with suitable PMC yield useful vital sign measurements without causing any discomfort to subjects.} \label{fig::covid_uav}
\end{figure}

    Remote sensing UAVs with multi-sensor modality are useful as telemedicine platforms. In a pandemic situation such as novel coronavirus (COVID-19), UAV radars equip the frontline medical professionals to perform their job with appropriate social distancing guidelines with minimum contact. 
    A highly autonomous UAV installed with a thermal imager, RGB depth camera, and radar sensor is capable of providing remote data collection and monitoring of patients in quarantine (Fig.~\ref{fig::covid_uav}). Health workers and medical professionals need data to manage a rapidly spreading respiratory pandemic. This UAV architecture is suitable for extensive instrument ion in the Internet-of-Things connectivity, where it communicates with and acquire patient information from a wearable sensor network. 
    In real-time, this aids in efficiently covering a large area in a short time for sanitization, thermal image collection, and patient monitoring without increasing the risks of the viral exposure to medical professionals.



\section{Summary}
\label{sec:summ}
The UAV radar platforms play a critical role in SRO by detecting living human-beings when they are occluded with debris or stranded in unfavorable locations. In LoS case, the radar VSD sensing complements the information obtained from existing cameras/sensors. 
The aerodynamic design, placement of radar, flight time and payload capabilities are important design considerations in UAV radar-based VSD applications. The simplest deployment mode exploits radar interferometry, wherein repeated scans of the human subject are used to extract micro-Doppler activity of body parts. Through-material imaging with UAV radar extracts VSD of humans not directly visible and beyond a wall. The recent works on combining SAR with VSD are helpful to clearly distinguish living subjects from the strong static clutter. Future directions include VSD with UAV swarms and tUAVs, developing advanced PMC techniques, and investigation for telemedicine during a pandemic.

  
\balance
\bibliographystyle{IEEEtran}
\bibliography{main}

\begin{thebibliography}{10}
\providecommand{\url}[1]{#1}
\csname url@samestyle\endcsname
\providecommand{\newblock}{\relax}
\providecommand{\bibinfo}[2]{#2}
\providecommand{\BIBentrySTDinterwordspacing}{\spaceskip=0pt\relax}
\providecommand{\BIBentryALTinterwordstretchfactor}{4}
\providecommand{\BIBentryALTinterwordspacing}{\spaceskip=\fontdimen2\font plus
\BIBentryALTinterwordstretchfactor\fontdimen3\font minus
  \fontdimen4\font\relax}
\providecommand{\BIBforeignlanguage}[2]{{%
\expandafter\ifx\csname l@#1\endcsname\relax
\typeout{** WARNING: IEEEtran.bst: No hyphenation pattern has been}%
\typeout{** loaded for the language `#1'. Using the pattern for}%
\typeout{** the default language instead.}%
\else
\language=\csname l@#1\endcsname
\fi
#2}}
\providecommand{\BIBdecl}{\relax}
\BIBdecl

\bibitem{zhao2019uav}
N.~Zhao, W.~Lu, M.~Sheng, Y.~Chen, J.~Tang, F.~R. Yu, and K.-K. Wong,
  ``{UAV}-assisted emergency networks in disasters,'' \emph{IEEE Wireless
  Communications}, vol.~26, no.~1, pp. 45--51, 2019.

\bibitem{erdelj2017help}
M.~Erdelj, E.~Natalizio, K.~R. Chowdhury, and I.~F. Akyildiz, ``Help from the
  sky: {L}everaging {UAV}s for disaster management,'' \emph{IEEE Pervasive
  Computing}, vol.~16, no.~1, pp. 24--32, 2017.

\bibitem{al2018remote}
A.~Al-Naji, A.~G. Perera, and J.~Chahl, ``Remote measurement of cardiopulmonary
  signal using an unmanned aerial vehicle,'' in \emph{IOP Conference Series:
  Materials Science and Engineering}, 2018, p. 012001.

\bibitem{perera2020detection}
A.~G. Perera, F.-T.-Z. Khanam, A.~Al-Naji, J.~Chahl \emph{et~al.}, ``Detection
  and localisation of life signs from the air using image registration and
  spatio-temporal filtering,'' \emph{Remote Sensing}, vol.~12, no.~3, p. 577,
  2020.

\bibitem{horton2019doctor}
T.~W. Horton, N.~Hauser, S.~Cassel, K.~F. Klaus, T.~Fettermann, and N.~Key,
  ``Doctor drone: {N}on-invasive measurement of humpback whale vital signs
  using unoccupied aerial system infrared thermography,'' \emph{Frontiers in
  Marine Science}, vol.~6, p. 466, 2019.

\bibitem{yan2018vital}
J.~Yan, Z.~Peng, H.~Hong, H.~Chu, X.~Zhu, and C.~Li, ``Vital-{SAR}-imaging with
  a drone-based hybrid radar system,'' \emph{IEEE Transactions on Microwave
  Theory and Techniques}, vol.~66, no.~12, pp. 5852--5862, 2018.

\bibitem{rong2021UAV}
Y.~Rong and D.~W. Bliss, ``Cardiac and respiratory sensing from a hovering
  {UAV} radar platform,'' in \emph{IEEE Statistical Signal Processing
  Workshop}, 2021, in press.

\bibitem{lin1975noninvasive}
J.~C. Lin, ``Noninvasive microwave measurement of respiration,''
  \emph{Proceedings of the IEEE}, vol.~63, no.~10, pp. 1530--1530, 1975.

\bibitem{rong2019smart}
Y.~Rong and D.~W. Bliss, ``Smart homes: {S}ee multiple heartbeats through wall
  using wireless signals,'' in \emph{IEEE Radar Conference}, 2019, pp. 1--6.

\bibitem{rong2019radar}
------, ``Is radar cardiography ({RCG}) possible?'' in \emph{IEEE Radar
  Conference}, 2019, pp. 1--6.

\bibitem{feng2014motion}
L.~Feng, L.-M. Po, X.~Xu, Y.~Li, and R.~Ma, ``Motion-resistant remote imaging
  photoplethysmography based on the optical properties of skin,'' \emph{IEEE
  Transactions on Circuits and Systems for Video Technology}, vol.~25, no.~5,
  pp. 879--891, 2014.

\bibitem{rong2020DepthSensor}
Y.~Rong and D.~W. Bliss, ``Respiration and cardiac activity sensing using {3-D}
  camera,'' in \emph{Asilomar Conference on Signals, Systems, and Computers},
  2020, in press.

\bibitem{camara2014cavalry}
D.~C{\^a}mara, ``Cavalry to the rescue: {D}rones fleet to help rescuers
  operations over disasters scenarios,'' in \emph{IEEE Conference on Antenna
  Measurements \& Applications}, 2014, pp. 1--4.

\bibitem{kishk2020aerial}
M.~{Kishk}, A.~{Bader}, and M.~S. {Alouini}, ``Aerial base station deployment
  in {6G} cellular networks using tethered drones: {T}he mobility and endurance
  tradeoff,'' \emph{IEEE Vehicular Technology Magazine}, vol.~15, no.~4, pp.
  103--111, 2020.

\bibitem{alaee2019radar}
M.~Alaee-Kerahroodi, K.~V. Mishra, and M.~R.~B. Shankar, ``Radar beampattern
  design for a drone swarm,'' in \emph{Asilomar Conference on Signals, Systems,
  and Computers}, 2019, pp. 1416--1421.

\end{thebibliography}

\end{document}